\documentclass[10pt,letterpaper,floatfix,showpacs,aps,prl,twocolumn]{revtex4}

\usepackage{latexsym,eucal,amsmath,amssymb,amsfonts,mathbbol,graphicx,color}
\usepackage{dcolumn}
\usepackage{bm}
\usepackage{amsmath}

\newcommand{\ket}[1]{|{#1}\rangle}
\newcommand{\bra}[1]{\langle {#1}|}
\newcommand{\sket}[1]{|{#1}\rangle\rangle}
\newcommand{\sbra}[1]{\langle\langle {#1}|}
\newcommand{\sbraket}[2]{\langle\langle{#1}|{#2}\rangle\rangle}
\newcommand{\sexp}[1]{\langle\langle {#1} \rangle\rangle}

\begin{document}

\title{Convergence rates for arbitrary statistical moments of
random quantum circuits}

\author{Winton G. Brown and Lorenza Viola}

\affiliation{\mbox{Department of Physics and Astronomy, Dartmouth
College, 6127 Wilder Laboratory, Hanover, NH 03755, USA}}

\begin{abstract}
We consider a class of random quantum circuits where at each step a
gate from a universal set is applied to a random pair of qubits, and
determine how quickly averages of arbitrary finite-degree polynomials
in the matrix elements of the resulting unitary converge to Haar
measure averages.  This is accomplished by establishing an exact
mapping between the superoperator that describes $t$-order moments on
$n$ qubits and a multilevel $SU(4^t)$ Lipkin-Meshkov-Glick
Hamiltonian.  For arbitrary fixed $t$, we find that the spectral gap
scales as $1/n$ in the thermodynamic limit.  Our results imply that
random quantum circuits yield an efficient implementation of
$\epsilon$-approximate unitary $t$-designs.
\end{abstract}


\maketitle


Random quantum states and unitary operators are broadly useful across
theoretical physics and applied mathematics.  Within quantum
information science \cite{NC}, they play a key role in tasks ranging
from quantum data hiding \cite{Terhal} and quantum cryptography
\cite{HHL} to noise estimation in open quantum systems
\cite{RM,ELL,EmersonUTD}. Unfortunately, generating an ensemble of
$N$-dimensional unitary matrices which are evenly distributed
according to the invariant Haar measure on $U(N)$ is inefficient, in
the sense that the number of required quantum gates grows
exponentially with the number of qubits, $n=\log_2 N$ \cite{NC}.
So-called {\em unitary $t$-designs} provide a powerful substitute for
Haar-distributed ensembles.  Building on the notion of a state
$t$-design \cite{EmersonTD}, a unitary $t$-design is an ensemble of
unitaries whose statistical moments up to order $t$ equal (exactly or
approximately) Haar-induced values \cite{EmersonUTD}.  That is, a
unitary $t$-design faithfully simulates the Haar measure with respect
to any test that uses at most $t$ copies of a selected $n$-qubit
unitary.  Ramifications of the theory of $t$-designs \cite{Eisert} are
being uncovered in problems as different as black hole evaporation and
fast ``scrambling'' of information \cite{Patrick}, efficient quantum
tomography and randomized gate benchmarking \cite{Paz}, quantum
channel capacity \cite{Matt}, and the foundations of quantum
statistical mechanics \cite{Low}.

Prompted by the above advances, significant effort has been devoted
recently to identifying {\em efficient} constructions of $t$-designs
and characterizing their convergence properties
\cite{Terhal,EmersonUTD,ODP,Znidaric,WintonPR,Harrow}.  Harrow and Low
established, in particular, the equivalence between approximate
$2$-designs and {\em random quantum circuits} as introduced in
\cite{RM}, and conjectured that a random circuit consisting of
$k$=poly$(n,t)$ gates from a two-qubit universal gate set yields an
approximate $t$-design \cite{Harrow}.  While supporting numerical
evidence was gathered in \cite{Braun} for low-order moments, and
efficient constructions of $t$-designs were reported in \cite{Harrow}
for any $t={\cal O}(n/\log n)$, the extent to which random quantum
circuits could be used to implement an approximate $t$-design for {\em
arbitrary, fixed} $t$ remained open.

In this Letter, we address this question by determining the rate at
which, for sufficiently large circuit depth, statistical moments of
arbitrary order converge to their limiting Haar values.  Our strategy
involves two steps: first, for given $t$, we show that the asymptotic
convergence rate is determined by the spectral gap of a certain
superoperator, which encapsulates moments up to order $t$; next, we
compute this gap by mapping the $t$-moment superoperator to a
multilevel version of the Lipkin-Meshkov-Glick (LMG) model, which is
known to be exactly solvable, and whose low-energy spectrum is well
understood in the thermodynamic limit $n \rightarrow \infty$
\cite{Vidal}.  Our approach ties together $t$-design theory with
established mean-field techniques from many-body physics, extending
earlier results by Znidaric \cite{Znidaric} for $t=2$.  Furthermore,
asymptotic convergence rates allow us to upper bound the convergence
time (minimum circuit length, $k_c$) needed for a desired accuracy
$\epsilon$ relative to the Haar measure to be reached.  For any fixed
$t$, we find that the scaling $k_c \sim n \log (1/\epsilon)$ holds for
sufficiently large $n$ and small $\epsilon$.


{\em Moment superoperator.$-$} Let a random quantum circuit of length
$k$ be a sequence $U_k \ldots U_1$ of $k$ unitary operators on an
$n$-qubit Hilbert space $\mathcal{H}=\otimes_j^n \mathcal{H}_{q_j}$,
where each $U_i$ is selected from an ensemble $\{\mu(U),U\}$, for a
probability distribution $\mu$ with support on a universal gate set.
To analyze arbitrary $t$-order moments, we introduce a Hilbert space
$\mathcal{H}_{M_t}=\mathcal{H}^{\otimes 2t}$, which consists of $2t$
copies of $\mathcal{H}$ and we refer to as the {\em moment space},
with $\dim(\mathcal{H}_{M_t}) \equiv D= N^{2t}$, and a {\em local
moment space} $\mathcal{H}_{l_t}$, which results from grouping factors
corresponding to the same qubit in $\mathcal{H}_{M_t}$.  That is,
$\mathcal{H}_{M_t} = \otimes_j^n \mathcal{H}_{q_j}^{\otimes 2t}=
\mathcal{H}_{l_t}^{\otimes n}$, with $\dim(\mathcal{H}_{l_t}) \equiv
d= 4^t$.  Moments of order $t$ may be described in terms of the
following linear operator on $\mathcal{H}_{M_t}$:
$$ M_t[\mu] = \int d\mu(U) U^{\otimes t} \otimes U^{* \otimes
t } \equiv \int d\mu(U) U^{\otimes t,t}.$$

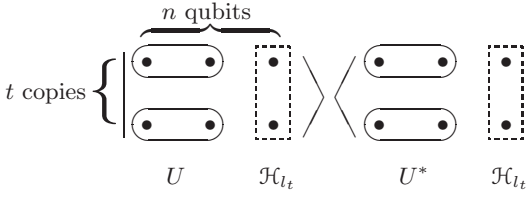
\begin{figure}[th]
\begin{center}
\begin{picture}(250,70)(-85,40)

\setlength{\unitlength}{.8pt}
\multiput(-20,65)(30,0){3}{$\bullet$}
\multiput(-20,95)(30,0){3}{$\bullet$}

\multiput(90,65)(30,0){3}{$\bullet$}
\multiput(90,95)(30,0){3}{$\bullet$}

\put(-10,118){\mbox{$n$ qubits}}
\put(-20,108){$\overbrace{~~~~~~~~~~~~~~~~~}$}
\put(-45,76){{\Huge\{}}
\put(-84,80){\mbox{$t$ copies}}

\put(36,40){$\mathcal{H}_{l_t}$}
\put(146,40){$\mathcal{H}_{l_t}$}

\put(-28,62){\line(0,1){40}}
\put(166,62){\line(0,1){40}}
\put(57,62){\line(1,2){9.5}}
\put(57,100){\line(1,-2){9.5}}
\put(81.5,62){\line(-1,2){9.5}}
\put(81.5,100){\line(-1,-2){9.5}}
\put(34.5,61){\dashbox{2}(16,44)}
\put(144.5,61){\dashbox{2}(16,44)}

\put(-2.5,68){\oval(43,15)}
\put(-2.5,98){\oval(43,15)}
\put(107.5,68){\oval(43,15)}
\put(107.5,98){\oval(43,15)}

\put(-8,40){$U$}
\put(102,40){$U^*$}

\end{picture}
\end{center}
\vspace*{-3mm}
\caption{The moment space $\mathcal{H}_{M_t}$ may be visualized as an
array of $2nt$ qubits, in such a way that $t$ copies support a ket in
the state space on $nt$ qubits, and the remaining $t$ copies the
corresponding bra.  In this way, a unitary $U$ on $n$ qubits induces a
transformation $U^{\otimes t,t}$ on density operators on $nt$
qubits. Dashed rectangles indicate the $2t$ qubits corresponding to a
local moment space $\mathcal{H}_{l_t}$, whereas the ovals correspond
to a unitary $U$ acting non-trivially on the first two qubits.}
\label{diagram}
\end{figure}

Physically, $M_t[\mu]$ may be viewed as the {\em superoperator}
induced by the action of $t$ copies of the random circuit on
$nt$-qubit density operators defined on $\mathcal{H}^{\otimes t}$ (see
Fig. 1).  In line with standard practice in open-system theory
\cite{Alicki}, we shall introduce ``operator kets'' in
$\mathcal{H}_{M_t}$, denoted $|A\rangle\rangle \equiv A$, and,
correspondingly, $\langle\langle A|=A^\dagger$.  Thus, a
$D^2$-dimensional operator ket transforms according to
$U^{\otimes t,t} |A\rangle \rangle \equiv UAU^{\dagger}$, under $U\in
U(N)$.  Once a basis for $\mathcal{H}_{M_t}$ is chosen, the matrix
representation of $M_t[\mu]$ specifies a complete set of $t$-order
moments.

The probability distribution that describes a random circuit of length
$k$, $\mu_k(U)$, is given by the $k$-th fold convolution of $\mu$ with
itself \cite{ELL,Harrow}.  That is, $ \mu_k(U)=\int \prod_{i=1}^k
d\mu(U_i) \delta(U-\Pi_{i=1}^k U_i).$ It then follows that
\begin{eqnarray*}
M_t[\mu_k]& = \hspace*{-1mm}&\int \Pi_{i=1}^k d\mu(U_i)\, \Pi_{i=1}^k
U_i^{\otimes t,t} \label{product} \\
& = \hspace*{-1mm}& \prod_{i=1}^k \int d\mu(U_i)\, U_i^{\otimes t,t}
=\big{(}M_t[\mu]\big{)}^k \equiv M_t^k[\mu].
\nonumber
\end{eqnarray*}
Note that, under the assumption that the ensemble $\{\mu(U),U\}$ is
invariant under Hermitian conjugation, $\mu(U)=\mu(U^\dagger)$,
$M_t[\mu]$ is an Hermitian operator on $\mathcal{H}_{M_t}$.

If $\mu(U)$ has support on a universal set of gates, then the measure
over the random circuit converges to the Haar measure on $U(N)$ in the
limit of infinite circuit length \cite{ELL}, $\lim_{k\rightarrow
\infty} \mu_k(U)=\mu_H(U)$. We begin by characterizing how these
convergence properties translate in terms of $t$-order moments. Let
$M_t[\mu_H] = \int d\mu_H(U) U^{\otimes t,t}$, and let
$$\mathcal{V}_t =\text{span}\{ |{\phi}\rangle\rangle \in
\mathcal{H}_{M_t} \, | \, U^{\otimes t,t} |{\phi}\rangle\rangle =
|{\phi}\rangle\rangle, \; \forall U\in {U}(N)\}$$
\noindent
be the subspace of fixed points of $U^{\otimes t,t}$, $U\in U(N)$,
with $\mathcal{P}_{\mathcal{V}_t}$ denoting the corresponding
projector.  We claim that
\begin{eqnarray}
\lim_{k\rightarrow\infty} \big{(}M_t[\mu]\big{)}^k =
\mathcal{P}_{\mathcal{V}_t} = M_t[\mu_{H}], \;\;\; \forall t.
\label{mtlimit}
\end{eqnarray}
While this is implied by the results
in \cite{Harrow}, a self-contained
proof follows.  Let $\sket{\phi}$ be an eigenoperator of $M_t[\mu]$
with eigenvalue $\lambda$, and $\ket{\phi_U}\rangle \equiv U^{\otimes
t,t} |{\phi}\rangle\rangle$.  Since
\begin{eqnarray*}
\big{|}\sexp{M_t[\mu]}\big{|} \hspace*{-0.8mm}=
\hspace*{-1mm} \Big{|}\int \hspace*{-.6mm}d\mu(U) \langle\langle{\phi} |
U^{\otimes t,t} |{\phi}\rangle\rangle \Big{|}
\hspace*{-0.8mm} \le \hspace*{-1mm} \int \hspace*{-.6mm}d\mu (U)
\big{|}\sbraket{\phi}{\phi_U}\big{|},
\end{eqnarray*}
it follows that $|\lambda| \le 1$, with equality holding if and only
if $U^{\otimes t,t} \sket{\phi}= \sket{\phi}$ for all $U$ with
$\mu(U)\ne 0$.  Any such operator ket $\sket{\phi}$ is also invariant
under any unitary of the form $U^{\otimes t,t}$, where $U$ is
generated by a random circuit of {\em arbitrary} length, that is,
$U=\Pi_{i=1}^k U_i$ for any $k$, as long as $\mu(U_i)\ne 0$.  Thus, if
$\mu(U)$ has support on a universal gate set, the eigenspace of
eigenvalue $1$ is precisely $\mathcal{V}_t$.  Since all other
eigenvalues of $M_t[\mu]$ have magnitude less than 1, $M_t^k[\mu]$
converges to $\mathcal{P}_{\mathcal{V}_t}$.  To establish the second
equality in Eq. (\ref{mtlimit}), we invoke the invariance of the Haar
measure under $U(N)$, $\mu_H(U)=\mu_H(U'U)$.  For $|{\phi}
\rangle\rangle $ an eigenoperator of $M_t[\mu_{H}]$ with eigenvalue
$\lambda$, it follows that, $M_t[\mu_{H}]|{\phi} \rangle\rangle =\int
d\mu_H(U) U^{\otimes t,t} |{\phi} \rangle\rangle =\lambda |{\phi}
\rangle\rangle $.  Thus,
\begin{eqnarray*}
&& U'^{\otimes t,t} \lambda |{\phi} \rangle\rangle = \int
\hspace*{-1mm}d\mu_H(U) (U'U)^{\otimes t,t} |{\phi} \rangle\rangle \\
&& =
\hspace*{-1.5mm}\int \hspace*{-1mm}d\mu_H(U'^\dagger U) U^{\otimes
t,t} |{\phi} \rangle\rangle = \hspace*{-1mm}\int
\hspace*{-1mm}d\mu_H(U) U^{\otimes t,t} \sket{\phi} = \lambda \sket{\phi}.
\end{eqnarray*}
If $\lambda\ne0$, it follows that $|{\phi} \rangle\rangle \in
\mathcal{V}_t$, otherwise $\lambda=0$, which establishes the desired
result.

Our next goal is to obtain the {\em rate} at which $M_t^k[\mu]$
approaches $M_t[\mu_{H}]$. Since $M_t[\mu_{H}]$ projects onto the
eigenspace of $M_t[\mu]$ of eigenvalue 1, the distance $\|
M_t^k[\mu]-M_t[\mu_{H}] \|$ with respect to any (unitarily invariant)
norm depends only on the remaining eigenvalues $\{\lambda_i\}$ of
$M_t[\mu]$ and the corresponding eigenprojectors
$\{\Pi_i\}$. Specifically, if $k$ is sufficiently large, $\|M_t^k[\mu]
- M_t[\mu_{H}]\|= \Big\|\sum_{\lambda_i\ne 1} \lambda_i^k\Pi_i\Big\|
\approx |\lambda_{1}|^k \|\Pi_1\|,$ where $\lambda_{1} \equiv
1-\Delta_t$ is the subdominant eigenvalue of $M_t[\mu]$.  Thus, the
asymptotic convergence rate is entirely determined by the spectral gap
$\Delta_t$ of $M_t[\mu]$.

{\em Mean-field solution.$-$} The starting point for mapping
$M_t[\mu]$ to an exactly solvable, infinitely-coordinated model is to
ensure that the following conditions are obeyed: (i) The applied
quantum gates consist only of single- and two- qubit gates selected
according to a distribution $\tilde{\mu}(U)$ on $U(4)$, with
$\tilde{\mu}(U)=\tilde{\mu}(U^\dag)$; (ii) The target pair of qubits
is picked uniformly at random.  We shall generally refer to the class
of circuits obeying (i)-(ii) as {\em permutationally invariant random
quantum circuits}.  Since, in each application of a random gate $U$ to
a fixed pair of qubits, the operator $U^{\otimes t,t}$ acts
non-trivially only on the associated bi-local moment space
$\mathcal{H}_{l_t}\otimes\mathcal{H}_{l_t}$, and this action is
identical for every qubit pair, the moment superoperator for any such
circuit may be written as follows:
\begin{equation}
 M_t[\mu]=\frac{2}{n(n-1)}\sum_{i<j=1}^n m_t^{ij}[\tilde{\mu}],
\label{H1}
\end{equation}
where for any pair $i,j$ the restriction $m_t[\tilde{\mu}]$ of
$m_t^{ij}[\tilde{\mu}]$ to $\mathcal{H}_{l_t}\otimes\mathcal{H}_{l_t}$
acts as $m_t [\tilde{\mu}]= \int \hspace*{-1mm}d\tilde{\mu}(U)
U^{\otimes t,t}$.  Recalling that dim($\mathcal{H}_{l_t})= d$,
$M_t[\mu]$ thus defines a qu$d$it Hamiltonian, which is invariant
under the symmetric group $\mathcal{S}_n$ of permutations of the $n$
local moment spaces.  Explicitly, if $\{b^i_{\alpha
\beta}=\ket{\alpha}\rangle \langle\bra{\beta} \}$ denotes an
outer-product basis for operators acting on any $\mathcal{H}_{l_t}$,
we may expand $m_t^{ij}=\sum_{\alpha\beta\gamma\delta=1}^{d}
\langle\langle \alpha \gamma | m_t | \beta \delta \rangle\rangle
b^i_{\alpha \beta} b^j_{\gamma \delta} \equiv
\sum_{\alpha\beta\gamma\delta=1}^{d} c_{\alpha\beta\gamma\delta}
b^i_{\alpha \beta} b^j_{\gamma \delta} $, and rewrite $M_t[\mu]$ as a
quadratic function of the collective operators $B_{\alpha
\beta}=\sum_{i=1}^n b_{\alpha \beta}^i$, that is,
$M_t[\mu]=\frac{1}{n(n-1)} \sum_{\alpha\beta\gamma\delta=1}^{d}
c_{\alpha\beta\gamma\delta} (B_{\alpha\beta}B_{\gamma\delta} -
\delta_{\beta\gamma} B_{\alpha\delta})$.
Since the operators $B_{\alpha \beta}$ obey $SU(d)$ commutation rules,
$[B_{\alpha \beta},B_{\gamma \delta}]= B_{\alpha \delta}\delta_{\beta
\gamma}-B_{\beta \gamma,}\delta_{\alpha \beta}$, $\mathcal{H}_{{M}_t}$
carries the (reducible) collective $n$-fold tensor product
representation of $SU(d)$, and $M_t[\mu]$ provides a $d$-level
extension of the standard, spin-$1/2$ LMG model \cite{Gilmore}.

Thanks to the invariance under $\mathcal{S}_n$, each of the
eigenoperators of $M_t[\mu]$ belongs to an irreducible representation
(irrep) of $SU(d)$.  Our first step is to show that the eigenspace
$\mathcal{V}_t$ of $M_t[\mu]$ corresponding to the ground-state
(extremal) eigenvalue of 1 lies in the totally symmetric irrep, of
dimension $d_S = \binom{4^t+n-1}{n}$~\cite{Elliott}.  Recall that
$\mathcal{V}_t$ consists of operators in $\mathcal{H}_{M_t}$ that
commute with all $t$-fold tensor power unitaries $U^{\otimes t}$. By
Schur-Weyl duality~\cite{Elliott,Harrow}, every such operator is a
linear combinations of elements of ${\mathcal S}_t$, under the natural
representation in $\mathcal{H}^{\otimes t}$. Note that the operators
spanning $\mathcal{V}_t$ are permutations of the $t$ copies of
$\mathcal{H}$ rather than permutations of the $n$ local moment spaces
$\mathcal{H}_{l_t}$. One may write any such permutation as
$|{\sigma^{(n)}} \rangle\rangle = \sum_{i_1\ldots i_t=1}^{N} \ket{
i_1\ldots i_t}\bra{i_{\sigma(1)} \ldots i_{\sigma(t)}},$ where $\sigma
\in \mathcal{S}_t$.
Furthermore, each such permutation may be viewed as a {\em product
ket} relative to the factorization $\mathcal{H}_{M_t}
=\mathcal{H}_{l_t}^{\otimes n}$.  Explicitly, $|{\sigma^{(n)}}
\rangle\rangle = \big(\ket{\sigma}\rangle\big{)}^{\otimes n}$, where
$\ket{\sigma}\rangle = \sum_{i_1\ldots i_t=0,1} \ket{i_1\ldots
i_t}\bra{i_{\sigma(1)} \ldots i_{\sigma(t)}}\in \mathcal{H}_{l_t}.$
The fact that $\mathcal{V}_t$ is {\em exactly spanned} by product
states is significant from the perspective of mean-field theory.  For
arbitrary $SU(d)$ quadratic Hamiltonians, it has been rigorously
established that the exact ground-state energy is given in the
thermodynamic limit by a mean-field Ansatz equivalent to assuming that
the ground state is an $SU(d)$ coherent state \cite{Gilmore}.  Since,
for the completely symmetric irrep, the manifold of coherent states
consists precisely of all product states \cite{Zhang}, the mean-field
extremal eigenspace of $M_t[\mu]$ is, in fact, {\em exact for any
n}.

The next step is to determine the lowest excitation energy in the
large-$n$ limit, which is accomplished by expanding $M_t[\mu]$ around
an arbitrary extremal mean-field state for each irrep
\cite{supplement}.  While, to our knowledge, a rigorous justification
of such a mean-field Ansatz is lacking, its validity for LMG
Hamiltonians is supported by an extensive body of theoretical and
numerical investigations \cite{Vidal}.  For the totally symmetric
irrep, the required diagonalization procedure is most
straightforwardly carried out by realizing the $U(d)$ algebra in terms
of $d$ canonical Schwinger boson operators $\{ a_\alpha,
a^\dagger_\beta\}$ \cite{okubo}.  That is, we let
$B_{\alpha\beta}=a^\dagger_\alpha a_\beta$ and rewrite the LMG
Hamiltonian as $M_t[\mu]=\frac{1}{n(n-1)}
\sum_{\alpha\beta\gamma\delta=1}^{d} c_{\alpha\beta\gamma\delta}
a^\dagger_\alpha a^\dagger_\gamma a_\beta a_\delta $.
Since the totally symmetric irrep of $\mathcal{H}_{M_t}$ contains
exactly $n$ Schwinger bosons, it is possible to eliminate one boson
mode by regarding it as ``frozen'' in the vacuum for a generalized
Holstein-Primakoff transformation \cite{okubo}.  Specifically, let the
local basis be chosen so that the frozen mode corresponds to
$\sket{\sigma}$, and let $\theta(n) \equiv
{(n-\sum_{\alpha\neq\sigma}a^\dagger_\alpha a_\alpha)^{1/2}}$, with
$a^\dagger_\sigma \rightarrow \theta (n)$, $a_\sigma \rightarrow
\theta (n)$. Two simplifications may now be invoked: first, the fact
that $\sket{\sigma^{(n)}}$ is an exact ground state causes any
coefficient of the form $c_{\alpha\sigma\beta\sigma}$,
$c_{\alpha\sigma\sigma\sigma}$ (and their complex conjugates) to
vanish; second, only terms up to the leading order in $1/n$ need to be
kept in $\theta(n)$. This finally yields:
$M_t[\mu]=1-\frac{1}{n} \sum_{\alpha \beta=1}^{d}
E_{\alpha\beta}a_\alpha^\dagger a_\beta + \mathcal{O}(1/n^2)$,
where $E_{\alpha\beta}=2(\delta_{\alpha\beta}-\sbra{\sigma
\alpha}m_t\sket{\sigma \beta}-\sbra{\sigma \alpha}m_t\sket{\beta
\sigma})$.  To leading order, the desired gap is then determined by
the smallest eigenvalue, $a_1$, of $E_{\alpha\beta}$.  That the latter
is {\em nonzero} may be shown by exploiting basic properties of the
superoperator $m_t$ \cite{supplement}.  This establishes our first
main result: For any permutationally invariant random quantum circuit,
and for any fixed $t>0$, the spectral gap may be expanded as
\begin{eqnarray}
\Delta_t = \sum_{p=1}^\infty a_p n^{-p} = \frac{a_1}{n} + {\mathcal O}
\Big(\frac{1}{n^2}\Big) ,
\label{gap}
\end{eqnarray}
for coefficients $\{a_p\}$ that may in general depend on $t$.

A stronger result may be obtained for a sub-class of random quantum
circuits which are, in addition, {\em locally invariant}, that is,
$\tilde{\mu}(U)$ is invariant under the subgroup $U(2)\times U(2)
\subset U(4)$ of local unitary transformations on the two target
qubits.  In this case, it is possible to choose a basis for each local
moment space $\mathcal{H}_{l_t}$, which includes a maximal set of
$t$-qubit operators $\{\sket{\omega}\}$ in the commutant of
$U^{\otimes t}$, with $U\in U(2)$.  Accordingly, every matrix element
$\sbra{\alpha\beta}m_t\sket{\gamma\delta}=0$, unless each local basis
element is itself an invariant, and the large-$n$ behavior of the gap
is determined by matrix elements of the form
$\sbra{\sigma\omega}m_t\sket{\sigma\omega}$ and
$\sbra{\sigma\omega}m_t\sket{\omega \sigma}$, with $\sigma\in
\mathcal{S}_t$ (without loss of generality, we may choose
$\sket{\omega}= \sket{I})$ and $\sket{\omega}$ an arbitrary
$U(2)$-invariant with $\sbraket{\sigma}{\omega}=0$.  Since, for $t>1$,
the maximum value of any such matrix element is independent of $t$
(see \cite{supplement} for full detail), it follows that the leading
order term $a_1$ {\em does not depend on $t$} for locally invariant
random quantum circuits.

\begin{figure}
\includegraphics[width=6cm]{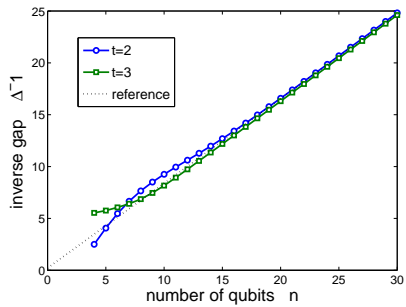}
\caption{\label{tgap} (Color online) Inverse spectral gap
$\Delta_t^{-1}$ of $M_t[\mu_H]$ with $t=2,3$ for a random circuit
consisting of two-qubit gates selected according to the Haar measure
on $U(4)$.
The line with slope $5/6$ corresponds to the asymptotic result. }
\end{figure}

{\em Example.$-$} Consider the simplest case where $t=2$ and
$\tilde{\mu}(U)=\mu_H(U)$ on $U(4)$.  The invariant eigenspace
$\mathcal{V}_2$ of $M_2$ is spanned by the identity
$\sket{I^{(n)}}=\big(\sket{I}\big{)}^{\otimes n}$ and the permutation
$\sket{S^{(n)}}=\big{(}\sket{S}\big{)}^{\otimes n}$ that swaps the
$t=2$ copies of $\mathcal{H}=\mathcal{H}_q^{\otimes n}$. Since
$\tilde{\mu}(U)$ is the Haar measure, $m_2$ coincides with the
projector onto the subspace $\mathcal{V}_2$ for $n=2$ qubits. An
orthogonal basis for $\mathcal{V}_2$ may be formed by taking even/odd
linear combinations under swap, $\sket{A_\pm}=
\sket{I^{(2)}}\pm\sket{S^{(2)}}$.  To find the excitation energies, we
choose one of the extremal local kets, $\ket{I}\rangle$, and minimize
$E_{\text{min}}=2\min
(1-\sbra{I\alpha}m_2\sket{I\alpha}-\sbra{I\alpha}m_2\sket{\alpha I}),$
over all local operators $\ket{\alpha}\rangle\in \mathcal{H}_{l_t}$
orthogonal to $\ket{I}\rangle$.  This yields
$\sket{\alpha}=\ket{S}\rangle-\langle\bra{S}I\rangle\rangle\ket{I}\rangle=
\sigma_1^1\sigma_1^2+\sigma_2^1\sigma_2^2+\sigma_3^1\sigma_3^2$,
and $\Delta_t=6/5n+{\mathcal{O}}(1/n^2)$. To determine how quickly the
large-$n$ scaling sets in, the fully symmetric sector of $M_t[\mu]$
under $\mathcal{S}_n$ was numerically diagonalized.  Since
${\mu}_H(U)$ is invariant under $U(2)\times U(2)$ transformations,
$\mathcal{H}_{l_2}$ may be restricted to the subspace of $SU(2)$
invariants.  From angular momentum theory \cite{Elliott}, the number
of such invariants is $\sum_J m_J^2=\frac{(2t)!}{(t+1)!t!}=C_t$, where
$m_J$ is the multiplicity of the $SU(2)$-irrep with total angular
momentum $J$.  This yields $d_S^{\text{loc}}= \binom{C_t+n-1}{n} \ll
d_S$, which makes numerical comparisons tractable for small $t$.
Exact results for $t$=2 and $3$ (see Fig. \ref{tgap}) indicate that
the scaling prediction for $\Delta_t$ becomes very accurate for $n
\gtrsim 14$.

{\em Convergence time.$-$} In order to establish the usefulness of a
random circuit as an $\epsilon$-approximate unitary $t$-design
\cite{EmersonUTD,Harrow}, we need to upper-bound the circuit length
required to achieve a specified accuracy, $\epsilon$.  Let the
convergence time with respect to a given norm be defined by the
minimum length $k_c$ for which $\|M_t^{k_c}[\mu]-M_t[\mu_{H}]\| \leq
\epsilon$.  That $M_t[\mu]$ be operationally indistinguishable from
$M_t[\mu_H]$ requires that the supremum of
$\|(M_t^{k}[\mu]-M_t[\mu_{H}])(\rho)\|_1$ be sufficiently small over
all $nt$-qubit density operators $\rho$.
We may bound the $1$-norm starting from the $2$-norm
\cite{Harrow}. For any density matrix $\rho$,
$\|(M_t^{k}[\mu]-M_t[\mu_{H}])(\rho)\|_2\le \lambda_1^k.$ This follows
from normalization of $\rho$ and the fact that $M_t[\mu_H]$ projects
onto the eigenspace of eigenvalue 1 of $M_t[\mu]$.  In conjunction
with the Cauchy-Schwartz inequality, this implies
$\|(M_t^{k}[\mu]-M_t[\mu_{H}])(\rho)\|_1\le 2^{nt}\lambda_1^k.$
Requiring that $2^{nt}\lambda_1^{k_c}\le\epsilon$ finally yields
$k_c\le\Delta_t^{-1}(\log(1/\epsilon)+nt\log(2))$. Since, using
Eq. (\ref{gap}), $\Delta_t^{-1} =\sum_{p=1}^\infty a_p'n^{2-p}\sim
a_1^{-1} n$ to leading order, $k_c = a_1^{-1} n \log(1/\epsilon)$ for
sufficiently small $\epsilon$.
It is worth stressing that we have {\em not} addressed how $k_c$
scales at fixed $n$ while letting $t\rightarrow \infty$. To answer
this question it is necessary to determine how $a_1'$ depends on $t$,
as well as the minimum value of $n$ required for the linear term
$a_1'n$ to dominate. While this is important for a full
characterization of $t$-designs, our results are directly relevant to
physical applications, where $t$ is fixed.

In summary, we have established that a large class of random quantum
circuits are efficient $\epsilon$-approximate unitary $t$-designs for
arbitrary finite $t$. The fact that the extremal eigenoperators are
separable suggests that similar results might be established for more
general random circuits for which the Hermiticity and the
$\mathcal{S}_n$-invariance assumptions of the moment superoperator
need not hold \cite{Long}.  As mentioned, a remaining open question is
to determine how the circuit length scales as the limits of large $n$
and large $t$ are taken together.  This may resolve the apparent
paradox that while Haar random unitaries are inefficient, arbitrary
$t$-designs are not, possibly with equal asymptotic rates.  We believe
that our findings, along with the techniques introduced to analyze
moments of unitary ensembles, will enable further understanding and
applications of $t$-designs across quantum physics.



\vspace*{-3mm}

\bibliographystyle{apsrev}

\newpage
\begin{appendix}
\begin{center}
{\bf \large Appendix: Supplementary Material}
\end{center}

\vspace*{1mm}

\begin{center}
{\bf Additional remarks on the determination of the spectral gap}
\end{center}

The diagonalization procedure illustrated in the main text determines
both the extremal and neighboring eigenvalues of $M_t$ belonging to
the totally symmetric irrep of $SU(d)$. Although expressed in terms of
bosonic operators, the procedure is equivalent to a variational Ansatz
whereby the trial wavefunction of the extremal state is of the form
$$\sket{\phi}^{\otimes n}=\sket{\phi\ldots\phi},$$
\noindent
and the first excited state, corresponding to a single bosonic
excitation, is of the form
$$ \ket{ n_\phi = n-1, n_\alpha\mbox{\small{=}}1}\rangle \equiv
\frac{1}{\sqrt{n}}(\sket{\phi \ldots\phi\alpha}+\ldots +\sket{\alpha
\phi \ldots \phi}).$$

In principle, it is possible that the subdominant eigenvalue may lie
instead in the $SU(d)$ irrep which carries exactly one anti-symmetric
pair of indexes (see {\em e.g.} Ref. [21]).  This can be accommodated
by using a different variational Ansatz for the excited state to be
minimized.  If, as in the main text, we choose a local basis in ${\cal
H}_{\ell_t}$ that includes the fixed extremal permutation
$|\sigma\rangle\rangle$, with the remaining local basis operators
$|\alpha \rangle\rangle \ne |\sigma\rangle\rangle$ treated as
excitation modes, the relevant single-excitation band is spanned by
kets of the following form:
$$\ket{ n_\sigma = n-1, n_\alpha\mbox{\small{=}}1}\rangle \equiv
\frac{1}{\sqrt{2}} \big{(}\ket{\sigma \ldots \sigma}\rangle\ket{\sigma
\alpha}\rangle-\ket{\alpha \sigma}\rangle\big{)},$$
\noindent
with nonzero matrix elements:
\begin{eqnarray*}
&&\langle\bra{n_\alpha=1}M_t\ket{n_\beta=1}\rangle=
\delta_{\alpha\beta} -
{\tilde{E}_{\alpha\beta}}/{n}+\mathcal{O}(1/n^2),\\ &&
\tilde{E}_{\alpha\beta} = 2(1-\langle\bra{\sigma \alpha} m_t
\ket{\sigma \beta}\rangle).
\end{eqnarray*}
(Note that the ``exchange term'' $\langle\bra{\sigma
\alpha}m_t\ket{\beta \sigma}\rangle$ is no longer present).
Upon diagonalization of ${\tilde{E}_{\alpha\beta}}$, the subdominant
eigenvalue $a_1$ is determined, with an identical $1/n$ scaling as
found for the symmetric irrep.
The possibility that the subdominant eigenvalue lies in this irrep may
be removed {\em a priori} by imposing a natural additional restriction
on the random quantum circuit, namely by requiring that the two-qubit
gate distribution be invariant under the transformation $S$ that swaps
the two-qubits, that is, $\tilde{\mu}(S U)=\tilde{\mu}(U
S)= \tilde{\mu}(U)$.

From a physical standpoint, it is also interesting to note that the
existence of {\em degenerate} mean-field ground states indicates that
$M_t[\mu]$ describes a deformed (or broken-symmetry) phase. For
generic $SU(d)$ models in the broken phase, the degeneracy of the
ground state manifold is known to be lifted at finite $n$ by tunneling
between the mean-field ground states, which induces corrections of
order $\exp(-an)$, and hence a gap that closes exponentially in the
thermodynamic limit, see {\em e.g.}, C. M. Newman and L. S. Schulman,
J. Math. Phys. {\bf 18}, 23 (1977); S. Dusel and J. Vidal,
Phys. Rev. B {\bf 71}, 224420 (2005) [Ref. [18] in the main text
provides additional relevant literature].  Remarkably, {\em no} such
correction can occur in our case because each extremal mean-field
state $\sket{\sigma^{(n)}}$ is an exact eigenoperator of $M_t[\mu]$
for arbitrary finite $n$, as stressed in the text.

\vspace*{5mm}

\begin{center}
{\bf Proof that the leading-order coefficient is non-vanishing}
\end{center}

We show here that, for the class of arbitrary permutationally
invariant random quantum circuits, the leading-order coefficient $a_1$
of the $1/n$ term in the spectral gap expansion is nonzero.

Recall that $a_1$ is given by the minimum eigenvalue of
$$E_{\alpha\beta}=2(1-\sbra{\sigma\alpha}m_t\sket{\sigma
\beta}-\sbra{\sigma \alpha}m_t\sket{\beta\sigma}),$$
\noindent
where $\ket{\alpha}\rangle$ and $\ket{\beta}\rangle$ are operators
orthogonal to a permutation
$\ket{\sigma}\rangle\in\mathcal{S}_t$. Thus, showing that $a_1>0$ is
equivalent to showing that $\sbra{\sigma \alpha}m_t\sket{\sigma
\alpha}+\sbra{\sigma \alpha}m_t\sket{\alpha \sigma}<1$, for any
operator $\ket{\alpha}\rangle$ such that
$\sbraket{\alpha}{\sigma}=0$. Let
$$\sket{\psi}=\frac{1}
{\sqrt{2}}(\sket{\sigma\alpha}+\sket{\alpha\sigma}).$$
\noindent
Upon taking the expectation value with respect to $m_t$, we have
\begin{eqnarray*}
\sbra{\psi}m_t\sket{\psi}= &\frac{1}{2}&
\Big(\sbra{\sigma\alpha}m_t\sket{\sigma \alpha}+\sbra{\sigma
\alpha}m_t\sket{\alpha \sigma} \\
&+&\sbra{\alpha\sigma}m_t\sket{\sigma
\alpha}+\sbra{\alpha\sigma}m_t\sket{\alpha \alpha}\Big).
\end{eqnarray*}
Since $M_t$ (and hence $m_t$) is invariant under interchange of any
two qubits, $\sbra{\alpha\sigma}m_t\sket{\sigma
\alpha}=\sbra{\sigma\alpha}m_t\sket{\alpha \sigma}$ and
$\sbra{\alpha\sigma}m_t\sket{\alpha\sigma}=
\sbra{\sigma\alpha}m_t\sket{\sigma\alpha}$, yielding
$$\sbra{\psi}m_t\sket{\psi}=\sbra{\sigma \alpha}m_t\sket{\sigma
\alpha}+\sbra{\sigma \alpha}m_t\sket{\alpha \sigma}.$$
\noindent
Following from the properties of the invariant subspace ${\mathcal
V}_t$ established in the main text, $\sbra{\psi}m_t\sket{\psi}=1$ if
and only if the operator $\ket{\psi}\rangle$ is invariant under
arbitrary unitary transformations of the form $U^{\otimes t}$ with
$U\in U(4)$, given that the corresponding two-qubit gate distribution
$\tilde{\mu}(U)$ is universal on $U(4)$.  Thus, we must show that
there exists a unitary transformation $U\in U(4)$ such that
$$U^{\otimes t} (\sigma\otimes \alpha + \alpha\otimes\sigma)
U^{\dagger \otimes t} \neq \sigma\otimes \alpha+ \alpha\otimes\sigma.$$
\noindent
(Recall that, in our notation, we identify $|\alpha\rangle\rangle
\equiv \alpha$, and so one).  In fact, we may take the permutation
$\ket{\sigma}\rangle$ to be the identity $\ket{I}\rangle$ without loss
of generality.  This follows upon noting that
$$\sbra{\psi}m_t\sket{\psi}=\int d\tilde{\mu}(U)
\mbox{tr}[\psi^\dagger U^{\otimes t}\psi U^{\dagger \otimes t}].$$
\noindent
Since $\sigma U^{\otimes t}=U^{\otimes t}\sigma$ for all $U\in U(4)$,
$\psi^\dagger\sigma U^{\otimes t}\sigma^\dagger\psi U^{\dagger \otimes
t}=\psi^\dagger U^{\otimes t}\psi U^{\dagger \otimes t}$.  Thus,
$\sigma\otimes\alpha+\alpha\otimes\sigma$ has the same expectation
value as $I\otimes\tilde{\alpha}+\tilde{\alpha}\otimes I$, where
$\tilde{\alpha}=\sigma^\dagger\alpha$.

Any operator $\ket{\tilde{\alpha}}\rangle$ orthogonal to the identity
may be expanded $\tilde{\alpha}=\sum_\nu c_\nu\sigma_{\nu_1}^1\ldots
\sigma_{\nu_t}^t$, where $\nu=(\nu_1,\ldots,\nu_t)$, and $\nu_i \in \{
0,1,2,3 \}$, with $\sigma_0=I$, $\sigma_1=\sigma_x$,
$\sigma_2=\sigma_y$, $\sigma_3=\sigma_z$, and the sum ranges over all
possible strings $\nu$ except the one where $\nu_i=0$~$\forall i$.
Thus, $\tilde{\alpha}\otimes I=\sum_\nu c_\nu \sigma_{\nu_1}^1\otimes
I^1 \ldots \sigma_{\nu_t}^t\otimes I^t$. Now, under
$U=\exp(\frac{\pi}{4}i\sigma_3\otimes\sigma_3)$, the following
transformations hold for Pauli operators:
\begin{eqnarray}
\sigma_1\otimes I &\mapsto &-\sigma_2 \otimes \sigma_3, \nonumber \\
\sigma_2\otimes I &\mapsto & \sigma_1\otimes \sigma_3,\nonumber\\
\sigma_3\otimes I & \mapsto &\sigma_3\otimes I, \nonumber\\ I\otimes
\sigma_1 & \mapsto &-\sigma_3 \otimes \sigma_2, \nonumber \\
I\otimes\sigma_2 &\mapsto & \sigma_3\otimes \sigma_1,\nonumber\\
I\otimes\sigma_3 & \mapsto &I \otimes \sigma_3. \nonumber
\label{trans0}
\end{eqnarray}
\noindent
Thus, for any $\sket{\tilde{\alpha}}$ with support on a Pauli string
such that $\nu_i = 1$ or $2$ for some $i$, there are terms in the
expansion of $U^{\otimes t} (I \otimes \tilde{\alpha} +
\tilde{\alpha}\otimes I) U^{\dagger \otimes t}$ which contain factors
of the form $\sigma_2 \otimes \sigma_3$, $\sigma_3 \otimes \sigma_2$,
$\sigma_1 \otimes \sigma_3$, and $\sigma_3 \otimes \sigma_1$. Since
there are no such term in the expansion of $I\otimes \tilde{\alpha} +
\tilde{\alpha}\otimes I $, it follows that $I\otimes \tilde{\alpha} +
\tilde{\alpha}\otimes I$ is {\em not} invariant under
$(\exp(\frac{\pi}{4}i\sigma_3\otimes \sigma_3))^{\otimes
t}$. Similarly, if $\sket{\tilde{\alpha}}$ has support on any Pauli
string such that $\nu_i = 3$ or $2$ for some $i$, then
$I\otimes\tilde{\alpha} + \tilde{\alpha}\otimes I$ is not invariant
under $(\exp(\frac{\pi}{4}i\sigma_1\otimes \sigma_1))^{\otimes t}$.
Since $\sket{\tilde{\alpha}}$ must belong to one of these two cases,
the proof is complete.\hfill$\Box$

\vspace*{5mm}

\begin{center}
{\bf Proof that the leading-order coefficient is $t$-independent}
\end{center}

We next show that, under the additional assumption that
$\tilde{\mu}(U)$ is invariant under the subgroup $U(2)\times U(2)
\subset U(4)$ of local unitary transformations on the two target
qubits, the leading order coefficient $a_1$ of the $1/n$ expansion
does not depend on $t$.

Recall that $a_1$ is determined by the maximum value of
$\sbra{\sigma\omega}m_t\sket{\sigma\omega}+
\sbra{\sigma\omega}m_t\sket{\omega\sigma}$, where $\sigma \in
\mathcal{S}_t$ and $\sket{\omega}$ is a $U(2)$ invariant which is
orthogonal to $|\sigma\rangle\rangle$. As shown previously, we may
take $\sket{\sigma}$ to be the identity without loss of generality.
Now, any operator orthogonal to the identity may be expanded as
$\omega=\sum_\nu c_\nu \sigma_{\nu_1}^{1}\ldots\sigma_{\nu_t}^{t}$
where, as before, $\nu=(\nu_1,\ldots,\nu_t)$, and $\nu_i \in \{
0,1,2,3 \}$, with $\sigma_0=I$, $\sigma_1=\sigma_x$,
$\sigma_2=\sigma_y$, $\sigma_3=\sigma_z$, and the sum ranges over all
possible strings $\nu$ except the one where $\nu_i=0$~$\forall i$.

The first step is to show that the expectation value
$\sbra{I\omega}U^{\otimes t,t}\sket{I\omega}$ for any $U\in U(4)$ may
be written as a symmetric polynomial of the form
\begin{equation}
\sbra{I\omega}U^{\otimes t,t} \sket{I\omega}=\sum_{2\le p_1+p_2+p_3 \le t}
 w_{\vec{p}} \,x^{p_1}y^{p_2}z^{p_3},
\label{pol}
\end{equation}
\noindent
where $x$, $y$, $z$ are real numbers in $[-1,1]$ which depend only on
$U$, and $w_{\vec{p}}$ are positive coefficients which depend only on
$\sket{\omega}$, with $\vec{p}=(p_1,p_2,p_3)$ and $p_i$ being
non-negative integers subject to $2\le p_1+p_2+p_3 \le t$.

To establish Eq. (\ref{pol}) above, we exploit the fact that an
arbitrary element of $U(4)$ may be written in the canonical
decomposition form $U=U_1\otimes U_2 \,U(q,r,s)\,U_1'\otimes U_2',$
where $U_1, U_1', U_2, U_2'$ act locally on either of two qubits and
$U(q,r,s)=\exp\{i(q \sigma_1\otimes\sigma_1+r
\sigma_2\otimes\sigma_2+s \sigma_3\otimes\sigma_3)\}$ [see {\em
e.g}. B. Kraus and J. I. Cirac, Phys. Rev. A {\bf 63}, 062309 (2001)].
Since $|\omega\rangle\rangle$ is a $U(2)$ invariant, it then suffices
to consider the action of $U(q,r,s)$.  Direct calculation shows that
under $U(q,r,s)$, the following transformations are obeyed by Pauli
operators:
\begin{eqnarray}
I \otimes \sigma_1 & \mapsto r_1s_1~ I \otimes \sigma_1 + r_2s_1~
\sigma_2\otimes \sigma_3 \nonumber \\ & -~ r_1s_2~ \sigma_3\otimes
\sigma_2 + r_2s_2~\sigma_1\otimes I, \label{trans1}\\ I \otimes
\sigma_2 & \mapsto s_1q_1~ I \otimes \sigma_2 + s_2q_1~
\sigma_3\otimes \sigma_1 \nonumber \\ & -~ s_1q_2~
\sigma_1\otimes\sigma_3 + s_2q_2~\sigma_2\otimes I, \label{trans2}\\ I
\otimes \sigma_3 & \mapsto q_1r_1~ I \otimes \sigma_3 + q_2r_1~
\sigma_1\otimes \sigma_2 \nonumber \\ & -~ q_1r_2~
\sigma_2\otimes\sigma_1 + q_2r_2~\sigma_3\otimes I,
\label{trans3}
\end{eqnarray}
\noindent
where $q_1=\cos(2q)$, $q_2=\sin(2q)$, and similar expressions hold for
$r_1,r_2$, and $s_1,s_2$, respectively.

The idea is now to evaluate $U^{\otimes t,t} |I\omega\rangle\rangle$
term by term in the expansion for $\sket{\omega}$, that is, we evaluate
\begin{eqnarray*}
U^{\otimes t} (I \otimes \omega )\,U^{\dagger \otimes t}
&\hspace*{-1mm}=\hspace*{-1mm}&
\hspace*{-1mm}\sum_\nu c_\nu U^{\otimes t}\,( I^{1} \otimes
\sigma_{\nu_1}^{1}\ldots I^{t}\otimes \sigma_{\nu_t}^{t})\, U^{\dagger
\otimes t} \\ \mbox{} &\hspace*{-1mm}=\hspace*{-1mm}& \sum_\nu c_\nu
\bigotimes_{i=1}^t U (I^{i} \otimes \sigma_{\nu_i}^{i}) U^\dagger,
\end{eqnarray*}
\noindent
where now the transformation rules in
Eq. (\ref{trans1})-(\ref{trans3}) may be applied to each of the $t$
factors independently.  Computing the matrix element $\sbra{I\omega}
U^{\otimes t,t} |I\omega\rangle\rangle$, there is a contribution of
the form $|c_\nu|^2 x^{p_1}y^{p_2}z^{p_3}$, arising from each of the
terms $c_\nu I^1\otimes \sigma_{\nu_1}^1\ldots
I^t\otimes\sigma_{\nu_t}^{t} $, where $x=r_1s_1$, $y=s_1q_1$,
$z=q_1r_1$, and $p_1$, $p_2$, $p_3$ are the number of instances where
$\nu_i=1, 2, 3$, respectively.  Summing over all terms in the
expansion for $\sket{\omega}$ finally results in a polynomial of the
form stipulated in Eq. (\ref{pol}), with $w_{\vec{p}}=
\sum_{\nu}|c_\nu|^2$ determined by the sum over all strings $\nu$ that
share the same $\vec{p}$-vector.  That the polynomial is symmetric
under the interchange of $x$, $y$, and $z$ follows from the invariance
of $\sket{\omega}$ under $U(2)$.  Note that by construction, $x$, $y$,
and $z$ are bounded between $[-1,1]$.

Let the degree of a Pauli string be the number of instances where
$\nu_i\ne 0$. Since, under $U^{\otimes t}$ with $U\in U(2)$, a Pauli
string can only be mapped to a Pauli string of equal degree, it
follows that any $U(2)$ invariant whose expansion contains terms of
differing degree can be written as a linear combination of $U(2)$
invariants each containing only terms of equal degree.  For $t=1$, the
only $U(2)$ invariant is the identity. For $t=2$, there is exactly one
$U(2)$ invariant orthogonal to the identity, namely,
$\omega=\frac{1}{\sqrt{3}}(\sigma_1^1\sigma_1^2+
\sigma_2^1\sigma_2^2+\sigma_3^1\sigma_3^2)$.  Consequently, no $U(2)$
invariant contains a term of degree 1, and the only degree-2 terms a
$U(2)$ invariant may contain are linear combinations of the form
$\sigma_1^i\sigma_1^j+\sigma_2^i\sigma_2^j+\sigma_3^i\sigma_3^j$.
Thus, for a monomial occurring in $\sum_{\vec{p}}
w_{\vec{p}}x^{p_1}y^{p_2}z^{p_3}$, $2\le p_1+p_2+p_2 \le t$, and if
$p_1+p_2+p_3=2$, then exactly one of $p_1$, $p_2$, or $p_3 =2$.

The next step to establish the claimed result is to show that
$\sum_{\vec{p}} w_{\vec{p}}x^{p_1}y^{p_2}z^{p_3} \le
\frac{1}{3}(x^2+y^2+z^2)$, where the right hand side is the polynomial
corresponding to $\sbra{I\omega_2}U^{\otimes t,t}\sket{I\omega_2}$,
where $\sket{\omega_2}$ is any degree-2 $U(2)$ invariant. To show this
we first show that the average over each set of monomials
$x^{p_1}y^{p_2}z^{p_3}$ defined by a set of integers $p\ge p'\ge p''$
distributed in every distinct way to $p_1$, $p_2$, and $p_3$, is less
than or equal to $\frac{1}{3}(x^2 + y^2 +z^2)$. There are two cases to
consider:

(i) If $p\ge 2$, then from $|x|, |y|, |z| \le 1$, it follows that
$x^{p}\frac{1}{2}(y^{p'}z^{p''}+ y^{p''}z^{p'}) \le x^2$,
$y^{p}\frac{1}{2}(z^{p'}x^{p''}+z^{p''}x^{p'}) \le y^2$, and
$z^{p}\frac{1}{2}(x^{p'}y^{p''}+ x^{p''}y^{p'}) \le z^2$. Thus, the
average of the left hand side of each inequality, which is the average
over the desired set of monomials, must be less than or equal to the
average of the right hand sides, which is $\frac{1}{3}(x^2+y^2+z^2)$.

(ii) If $p=p'=p''=1$, then using $x=r_1s_1$, $y=s_1q_1$ and
$z=q_1r_1$, $xyz\le \frac{1}{3}(x^2+y^2+z^2)$ can be written
$q_1^2r_1^2s_1^2 \le \frac{1}{3}(r_1^2s_1^2+s_1^2q_1^2+q_1^2r_1^2)$,
which follows from $q_1^2r_1^2s_1^2\le r_1^2s_1^2, s_1^2q_1^2,
q_1^2r_1^2$.

\noindent
Since $\sum_{\vec{p}} w_{\vec{p}} x^{p_1}y^{p_2}z^{p_3}$ is a weighted
average of the above monomial averages, each of which is less than or
equal to $\frac{1}{3}(x^2+y^2+z^2)$, it follows that $\sum_{\vec{p}}
w_{\vec{p}} x^{p_1}y^{p_2}z^{p_3}\le\frac{1}{3}(x^2+y^2+z^2)$.

Finally, the steps described above can be applied to the exchange term
$\sbra{I\omega}U^{\otimes t,t}\sket{\omega I}$, resulting in
$\sbra{I\omega }U^{\otimes t,t} \sket{\omega
I}\le\frac{1}{3}(x^2+y^2+z^2)$, where $x=b_2c_2$, $y=a_2c_2$,
$z=a_2b_2$ for any $U\in U(4)$.  Since these inequalities hold for
every $U\in U(4)$, it follows that
\begin{eqnarray*}
\sbra{I\omega}m_t\sket{I\omega}&+&\sbra{I\omega}m_t\sket{\omega I}\\
& \le &
\sbra{I\omega_2}m_t\sket{I\omega_2}+\sbra{I\omega_2}m_t\sket{\omega_2I},
\end{eqnarray*}
\noindent
where $\sket{\omega_2}$ is any degree-$2$ $U(2)$ invariant. Since
$\sbra{I\omega_2}U^{\otimes
t,t}\sket{I\omega_2}=\frac{1}{3}(x^2+y^2+z^2)$ (and the equivalent
expression for $\sbra{I\omega_2}U^{\otimes t,t}\sket{\omega_2 I}$)
holds for every degree-2 $U(2)$ invariant, and the expression does not
depend on $t$, it follows that the maximum of
$\sbra{I\omega}m_t\sket{I\omega}+\sbra{I\omega}m_t\sket{\omega I}$
over all $\sket{\omega}$ such that $\sbraket{\omega}{I}=0$ is given by
$\sbra{I\omega_2}m_2\sket{I\omega_2}+\sbra{I\omega_2}m_2\sket{\omega_2
I}$.  This concludes the proof.  \hfill$\Box$

\end{appendix}
\end{document}